# Study on the planar segregation of solute atoms in Mg-Al-Gd alloy

Xin-Fu Gu[1,2*], Tadashi Furuhara[2], Leng Chen[1], Ping Yang[1]

1. School of Materials Science and Engineering, University of Science and Technology Beijing, Beijing, 100083, China

2. Institute for Materials Research, Tohoku University, Sendai, 980-8577, Japan

*Corresponding author: Xin-Fu Gu, xinfugu@ustb.edu.cn

**Abstract**

Solute segregation plays an important role in forming long-period stacking ordered (LPSO) structure in Mg-M-RE (M: Zn, Ni etc., RE: Y, Gd, etc.) alloys. In this work, the planar segregation in Mg-Al-Gd alloy is characterized by high angle annular dark field (HAADF)-scanning transmission electron microscopy (STEM) and three-dimensional atom probe (3DAP). It is found that the planar segregation can be considered as Guinier-Preston (G.P.) zone, since there is no planar fault or ordered structure within the segregation. The G.P. zone mainly enriches with Gd atoms based on 3DAP data. Furthermore, the segregation behaviour is examined by first-principles calculation.

**Keywords:** Magnesium alloy, Segregation, HAADF-STEM, 3DAP, First-principles

Long-period stacking ordered (LPSO) structures are found to be a new promising strengthening phase for Mg-Zn-RE (RE: rare earth element) alloys, such as Mg-Zn-Y, Mg-Zn-Gd etc. [1-6]. In general, LPSO structure can also be found in Mg-M-RE (M: Zn, Cu, Ni, Al, or Co, RE: Y, Gd, Tb, Dy, Ho, Er, Tm) alloy systems [7-11]. The LPSO structure often consists of regular arrangements of four-layer-height fcc structural units (SUs) separated by several Mg layers (hcp structure) paralleling to $(0001)_{\alpha}$ plane of hcp ($\alpha$) matrix, and the fcc SUs have a ABCA type stacking sequence, and they usually enrich with RE and M elements [12, 13]. In recent studies, the ideal fcc SU is proposed to contain ordered $M_6RE_8$ clusters with $L1_2$ type structure [10, 13]. Commonly observed LPSO structures 10H, 18R, 14H, and 24R correspond to the intermediate Mg layers of 1, 2, 3 and 4 layers, respectively [14, 15], where the symbols for the LPSO structures are Ramsdell's notation which specifies the total number of $(0001)_\alpha$ layers in the hexagonal unit cell followed by the letter H or R to indicate the lattice type.

The transformation from hcp structure to fcc SU needs two processes. One is the structure change by operating a $<1\text{-}100>_\alpha/3$ type partial dislocation on $(0001)_\alpha$ plane to change the stacking order from hcp to fcc, and the other is enrichment of M and RE elements by diffusion process. This kind of transformation can be treated as displacive-diffusional transformation process [16-18]. The order of these two processes is crucial to understand the atomic mechanism during formation of LPSO structure, and the possible transformation sequences are summarized in Figure 1 in Ref. [19] by Okuda *et al.*. They proposed a hierarchical transformation mechanism based on their synchrotron radiation measurements in an Mg-Zn-Y alloy [20]. The solute cluster forms first, subsequently the clusters are spatially arranged accompanying the introduction of stacking faults. However, the stage of stacking faults formation is not directly observed. With the aid of high angle annular dark field (HAADF) scanning transmission electron microscopy (STEM), Kishida *et al.* [21] reported the planar segregation of solute atoms in Mg-Al-Gd alloy for the first time. However, the solute atoms Al and Gd are proposed to be co-segregated at this region without direct evidence. Recently, Kim *et al.* [18] found that the co-segregation of Zn and Y atoms forms Guinier-Preston (G.P.) zones in Mg-Zn-Y alloys by using HAADF-STEM and STEM-EDS, and the G.P. zone will transformed to SU by operating of partial dislocation. Matsunaga observed similar phenomenon in Mg-Zn-Gd alloy [22], but the chemical information of G.P. zone was not measured. Nevertheless, the planar segregation seems the precursor phase for formation of LPSO structure. Understanding of composition and structure of the segregation will be benefit to understand the transformation mechanism of LPSO structure. However, the reported chemical information of segregation is limited. First of all, the atomic species deduced from the Z-contrast method in HAADF-STEM mode should have large difference in atomic number [23], such as RE comparing with other elements in Mg-M-RE alloy, thus it is hard to differentiate two atoms with near atomic numbers. Secondly, the signal for Energy Dispersive X-ray Spectroscopy (EDS) is weak due to small interaction volume and image-drift is another problem during the measurement.

Three-dimensional atom probe (3DAP) tomography could provide three-dimensional maps of chemical heterogeneities in Mg alloy with nano-scale resolution, such as nano-scale precipitates in Mg alloy [24, 25] and clusters of solute atoms [26, 27]. To our knowledge, Inoue *et al.* [28] was the first to apply 3DAP to study the segregation accompanying the well-formed LPSO structures, i.e. at later stage growth of LPSO structure, in Mg-Zn-Y and Mg-Al-Gd alloys. They found only Zn enriched layer on top of 14H LPSO in Mg-Zn-Y alloy, while there is no segregation on top of 18R LPSO. One possible reason may be due to the dynamic feature of the segregation. In Mg-Al-Gd alloy, they conclude the segregation enriches with Gd but is poor at Al, though the Al peak can be clearly seen in their composition profile. Based on these facts, the solute atoms may be not synchronized during LPSO formation. In order to verify this proposal, the early stage of LPSO formation is studied. On one hand, the initial stage of chemical modulation can be studied. On the other hand, one may capture a large number of initial stages to catch the different status of segregation.

The LPSO structure in Mg-Al-Gd alloy shows well-ordered structure [10, 21], and this alloy system is a good candidate to investigate the possible cluster structure and composition in segregated area. Therefore, the segregation in Mg-Al-Gd alloy will be characterized by STEM and 3DAP techniques in this work, and interaction between solute atoms will be discussed by using first-principles calculation.

Both as-casted $Mg_{92}Al_3Gd_5$ alloy (at.% default) and isothermal treated specimen are used in this study. For isothermal treatment, the sample is held at 500°C for 4 hours. The method for preparing TEM (HAADF-STEM) specimens and observation condition are the same as in our previous work [29]. The 3DAP data was taken by CAMECA LEAP 4000HR at 35 K. A pulse fraction of 20% and a pulse frequency of 200 kHz were applied.

Figure 1(a-c) shows some typical atomic-resolution HAADF-STEM images along $[11\text{-}20]_\alpha$ and $[01\text{-}10]_\alpha$ zone axes of hcp (α) matrix in as-casted Mg-3Al-5Gd alloy. The bright contrasts in the images arise from the enrichment of Gd according to Z-contrast principle [23], because the atomic numbers (Z) for three elements in this study are 12 for Mg, 13 for Al and 64 for Gd, respectively. The segregation with four atomic layers height and fcc structure is fcc SU, such as in Figure 1(a, b). The fcc stacking sequence of SU as indicated as a yellow slash in the images. From the contrast profile shown in the right part of Figure 1(a), apart from the strong peak due to the SU, there are other two peaks (indicated by two headed arrows in the figure) at both top and bottom of the SU. These places are also enriched with solute atoms Gd. In corresponding HAADF-STEM image, they show planar-like segregation spanning several layers, but with hcp structure as in the matrix. Namely, the planar segregations are G.P. zones. There is a solute-deplete zone between the G.P. zone and SU. The solute-deplete zone across 2~3 $(0001)_\alpha$ layers. The G.P. zone can also be observed at the top 18R LPSO structure as shown in Figure 1(c). These results are consistent with the work done by Kishida *et al.* [21]. In order to check whether there are ordered cluster structures in G.P. zone, the $\langle 01\text{-}10 \rangle_\alpha$ zone axis is selected and the image is shown in Figure 1(b). Within the SU, well-defined clusters ($L1_2$ type structure [10, 13]) can be found at this zone axis, and a single cluster is indicated by a box in Figure 1(b). However, contrary to the model of planar segregation [20] and the calculation [30], no obvious ordered cluster is observed in the G.P. zone.

This point was further supported by the observation in the aged sample. Figure 1(d-e) shows the 18R LPSO structure after ageing for 4h at 500°C. These two images are taken at the same area, but with different zone axis. It is observed that well-ordered structure is formed in 18R LPSO structure during ageing. A partial transformed SU is indicated by a yellow arrow in Figure 1(d-e). The left part of the atomic rows has an hcp (α) structure while the right side has an fcc structure. As can be seen from Figure 1(e) at zone axis of $[01\text{-}10]_\alpha$, the transformed fcc structure has an ordered structure, but the segregated front is not. It seems the segregation will help the nucleation of planar defects, and the ordered structure only takes place after the formation of planar defects.

At the initial stage, the G.P. zone could also resemble the periodicity of SUs in the LPSO structure as shown in Figure 2(a-c). Several SUs could be observed in this figure, and G.P. zone are located between them. After the G.P. zone transformed to SU during ageing, it would form a mixed 18R and 14H LPSO, since the interspacing between the SUs is around 2~3 $(0001)_{\alpha}$ layers. Based on the Z-contrast image, it can be qualitatively deduced that the composition of solute atoms in G.P. zones should be lower than that of SU due to its weaker contrast.

The chemical information of segregated area is further quantitatively measured by 3DAP. Figure 3(a) shows the element distribution measured by 3DAP. The SU enriches both Gd and Al and can be identified. Apart from these contrasts for SU, there are some extra segregation in Gd element mapping as indicated by arrows. According to the 3D element distribution, both G.P. zones and SUs exhibit planar shape. Furthermore, the composition profile is generated based on the reconstructed element mapping as shown in Figure 3(b). Accordingly, Gd content for SU is 9.5~13 at.%, and Al content in SU is about 7-10 at.% (the Gd and Al content for ideal SU is 16.7% and 12.5%, respectively). However, for G.P. zone, Gd content varies from 5-6.5 at.%, while corresponding Al content varies from 1-3 at.%. It is noted that the Al content of about 1 at.% is nearly the same with bulk content in matrix, thus there should be no segregation of Al at this composition, such as the positions (indicated by the arrow) at the bottom of element mapping. Apparently, different stage of segregation for formation of SU is successfully captured. The formation process of LPSO/SU structure could be deduced as follows: Gd atoms firstly segregate, then Al atoms segregate, finally the stacking fault forms accompanying the ordering of Al and Gd to form an ordered structure. The finally ordered structure would be similar to Figure 3(c) after long-period ageing.

In order to understand the segregation behavior of solute atoms, the firs-principles calculation is carried out in this study. Since the G.P. zone appears together with SU, the interaction between solute atoms and the SU is considered similar to the study in Mg-Zn-Y system [31]. The atomic model is shown in Figure 4(a), where twenty layers containing two ordered SUs are modeled. The basal size is selected to be $\sqrt{3}a \times \sqrt{3}a$, where $a$ is the lattice parameter of matrix. During the calculation, the atoms in stacking layer A or C between two SUs are replaced by a solute atom Al or Gd, and the total energy is then calculated. Only layers 1~6 in Figure 4(a) are considered due to the mirror symmetry of this structure. It is noted that there are different inequivalent sites on each layer [31]. In our model, the inequivalent sites in layer A and C are shown in Figure 4(b) and (c), respectively. There are four inequivalent sites in layer A denoted by $A_1$, $A_2$ $A_3$ and $A_4$, while three inequivalent sites in layer C are denoted by $C_1$, $C_2$ and $C_3$， as used by Kishida *et al.*[10, 32]. The first-principles calculation was carried out with the software named Cambridge Serial Total Energy Package (CASTEP) [33]. Electron exchange correlation functional was described by the Perdew-Bruke-Ernzerhof (PBE) version of generalized gradient approximation (GGA) [34]. The tolerance for energy change was set to be $5.0 \times 10^{-6}$ eV/atom. The calculated results for different positions and layers are shown in Figure 4(d). As for the Gd atoms, the interaction energy with SU decreases with increasing separation distance. It means Gd atoms prefer to be far from the SU, and forms G.P. zones. This possibly explains why there is a solute depleted zone between the SU and G.P. Zone observed by HAADF-STEM. Al has a preferred site $C_2$ in Layer 1 due to the strong interaction between Al and Gd in the SU. At other positions, the interaction energy for Al does not vary too much as for Gd atoms. After Gd atoms segregated, Al will segregate in the same area, because Al-Gd has negative binding energy comparing to Al-Al and Gd-Gd pairs [30]. Moreover, Gd segregation can reduce the intrinsic stacking fault energy in $(0001)_\alpha$ plane [35], which can be helpful to form the fcc SU. With the planar defects, the ordered clusters could form as simulated by Kimizuka and Ogata [30].

In summary, the atomic structure and chemical composition of planar segregation in Mg-Al-Gd alloy is characterized by HAADF-STEM and three-dimensional atom probe, respectively. The planar segregation often forms together with LPSO structure or its structure units, and the periodic distribution of segregation resembles the periodicity between structure units in the LPSO structure. The planar segregation can be treated as G.P. Zone, since it has an hcp structure and does not contain obvious cluster structure. The G.P. Zone firstly enriches with Gd atoms, and then Al co-segregated due to negative binding energy. Finally, the ordered clusters would form accompanying the formation of the stacking fault in G.P. zone.

**Acknowledgement**

This work was supported by a Grant-in-Aid for Scientific Research on Innovative Areas, "Synchronized Long-Period Stacking Ordered Structure", from the Ministry of Education, Culture, Sports, Science and Technology, Japan (No.23109006) and Fundamental Research Funds for the Central Universities (No. FRF-TP-17-003A1). The Mg alloy used in this study was supplied by Kumamoto University. Special thanks to Mr. Y. Hayasaka for technical support at Electron Microscopy Center, Tohoku University, Japan.

Figures

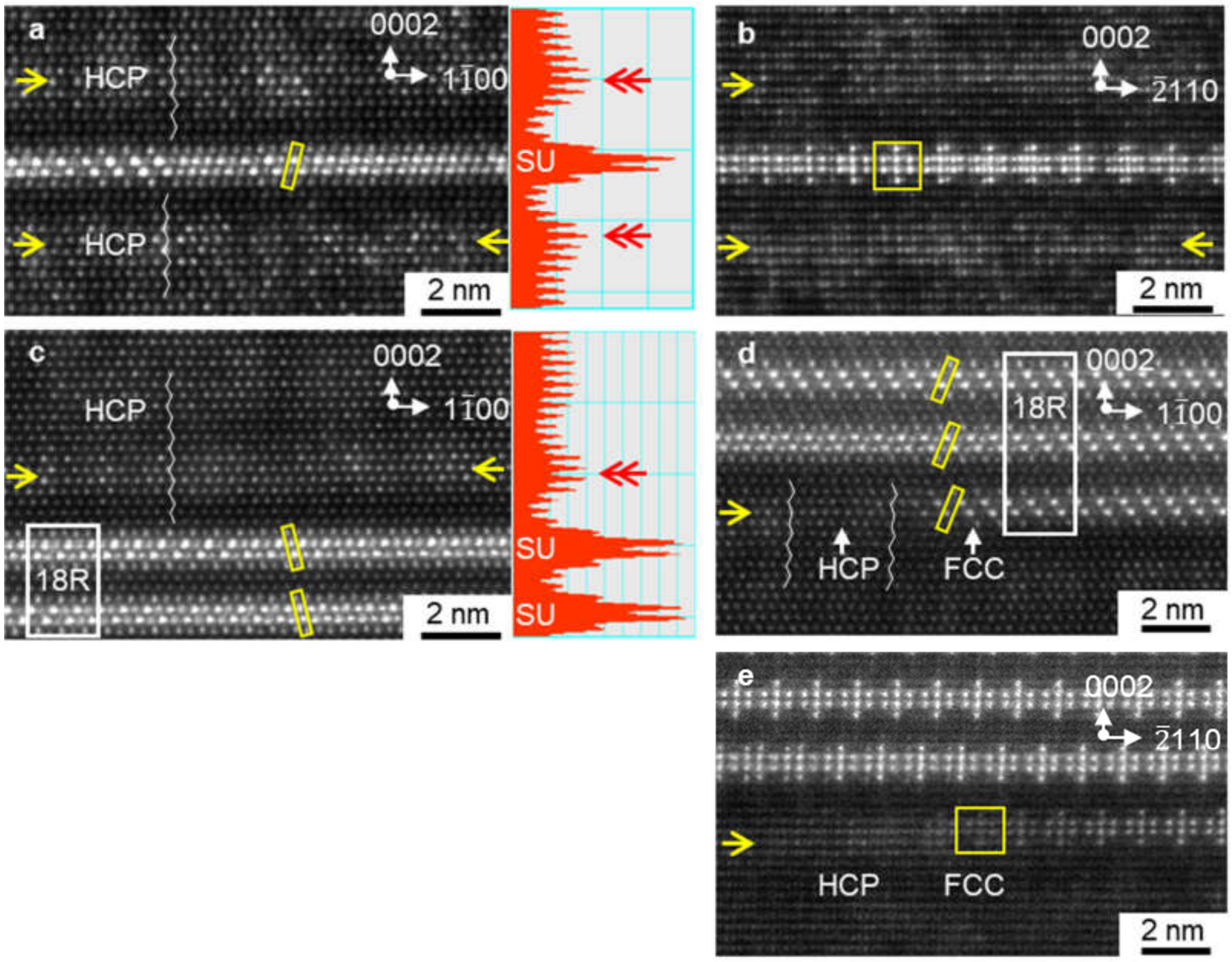

Figure 1. Planar segregation observed by HAADF-STEM. a) segregation around single fcc structure unit in as-casted alloy viewed along $[11\text{-}20]_{\alpha}$ of hcp (α) matrix, b) view of figure (a) from another direction $[01\text{-}10]_{\alpha}$ of hcp (α) matrix, c) segregation during growth of 18R LPSO structure in as-casted alloy viewed along $[11\text{-}20]_{\alpha}$, d) segregation during growth of 18R LPSO structure in the sample aged for 4h at 500°C viewed along $[11\text{-}20]_{\alpha}$. e) another zone axis view of figure d) from $[01\text{-}10]_{\alpha}$. The right part of (a) and (c) shows its corresponding contrast profile. The stacking sequence in fcc structure unit is indicated by a slash. The cluster in the structure units is marked by a yellow box. The arrows indicate the segregation in hcp matrix.

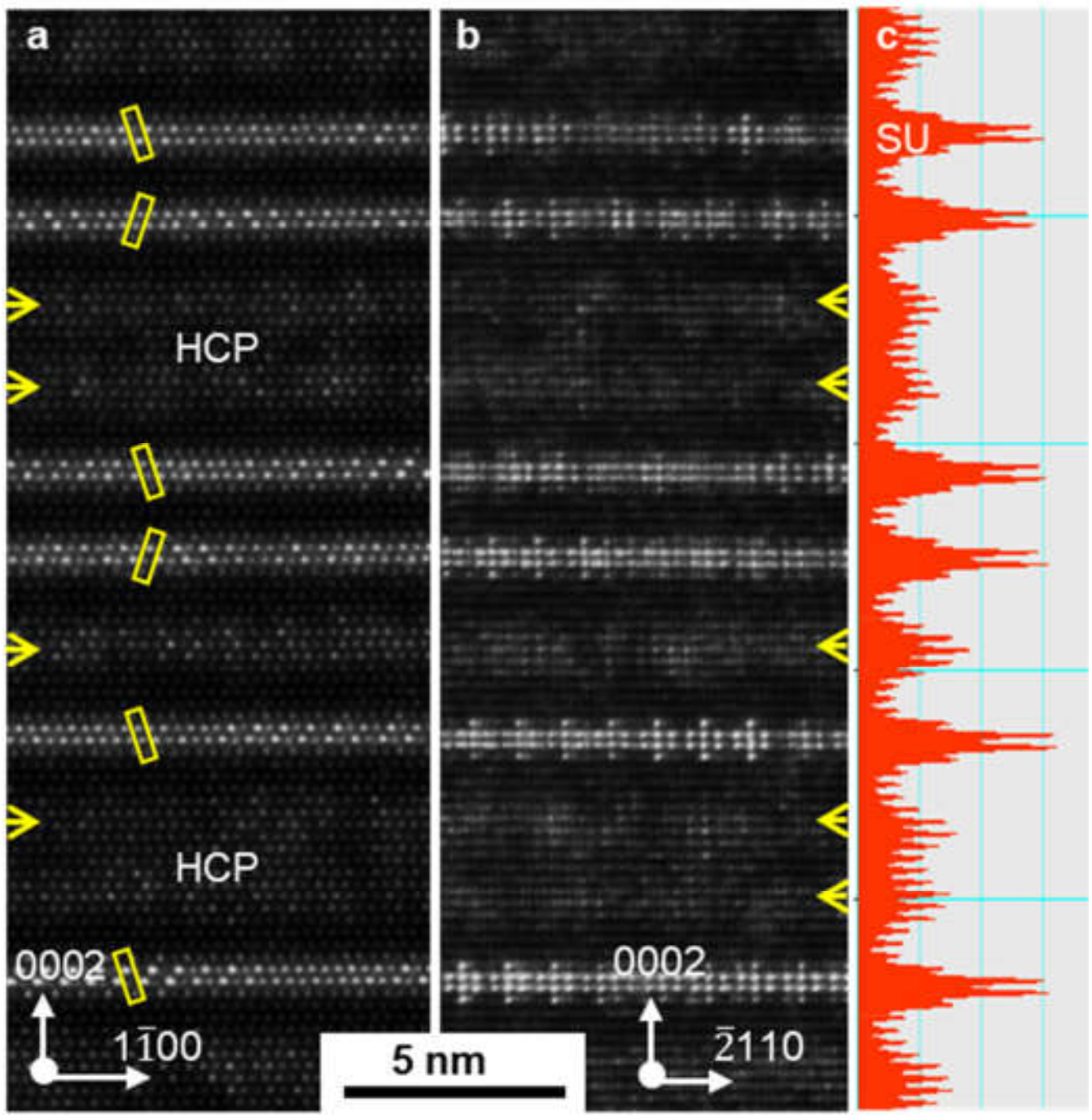

Figure 2. Distribution of G.P. zone in Mg-Al-Gd alloy. a) HAADF-STEM image of G.P. zone in as-casted state viewed along $[11\text{-}20]_\alpha$ zone axis of hcp ($\alpha$) matrix, b) corresponding image of figure a) viewed along $[01\text{-}10]_\alpha$ zone axis, c) contrast profile.

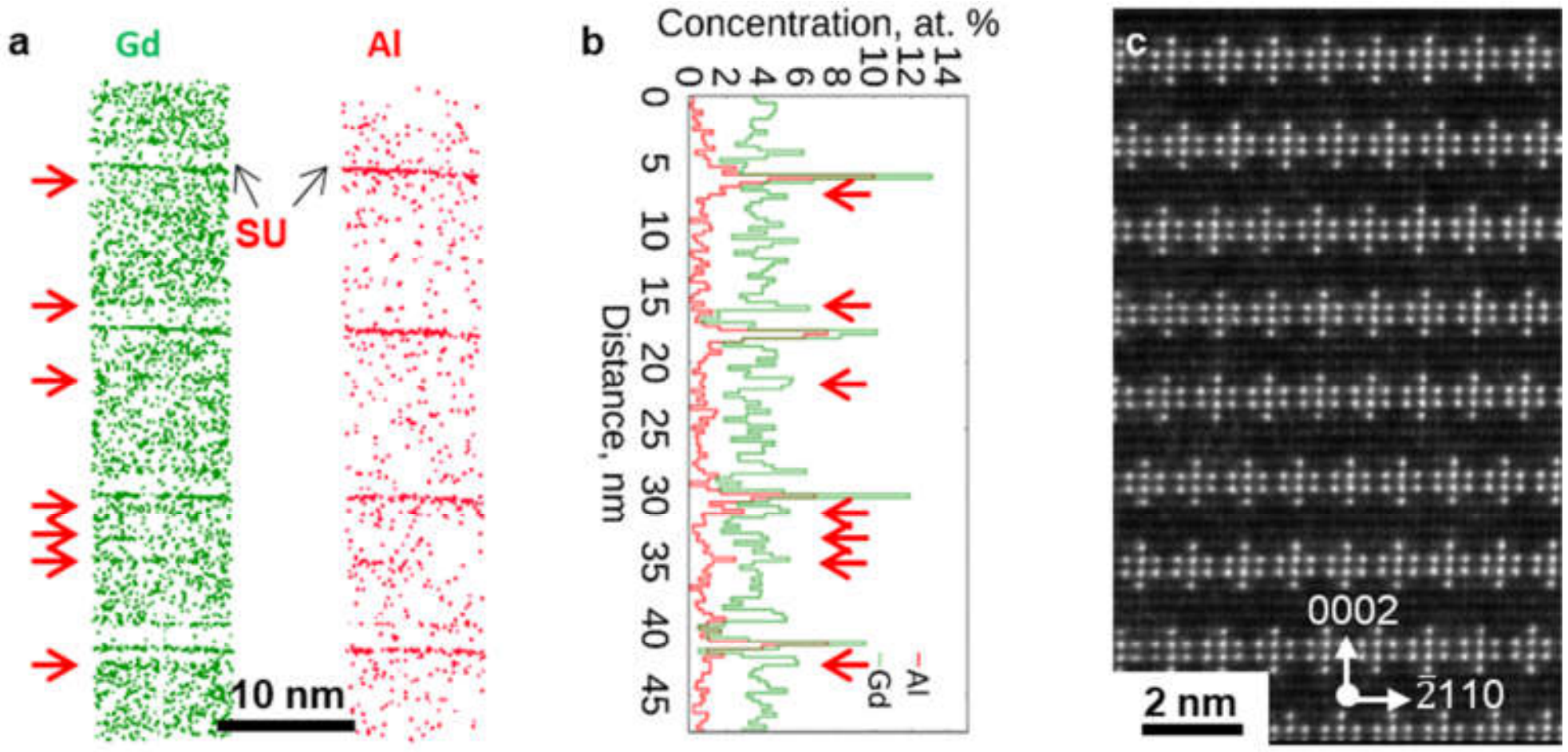

Figure 3. (a) Element mapping of Gd, Al by 3DAP, (b) composition profile along vertical axis, c) HAADF-STEM image of 18R LPSO structure after ageing 4h@500°C viewed along $[01\text{-}10]_{\alpha}$. G.P. Zones are indicated by arrows.

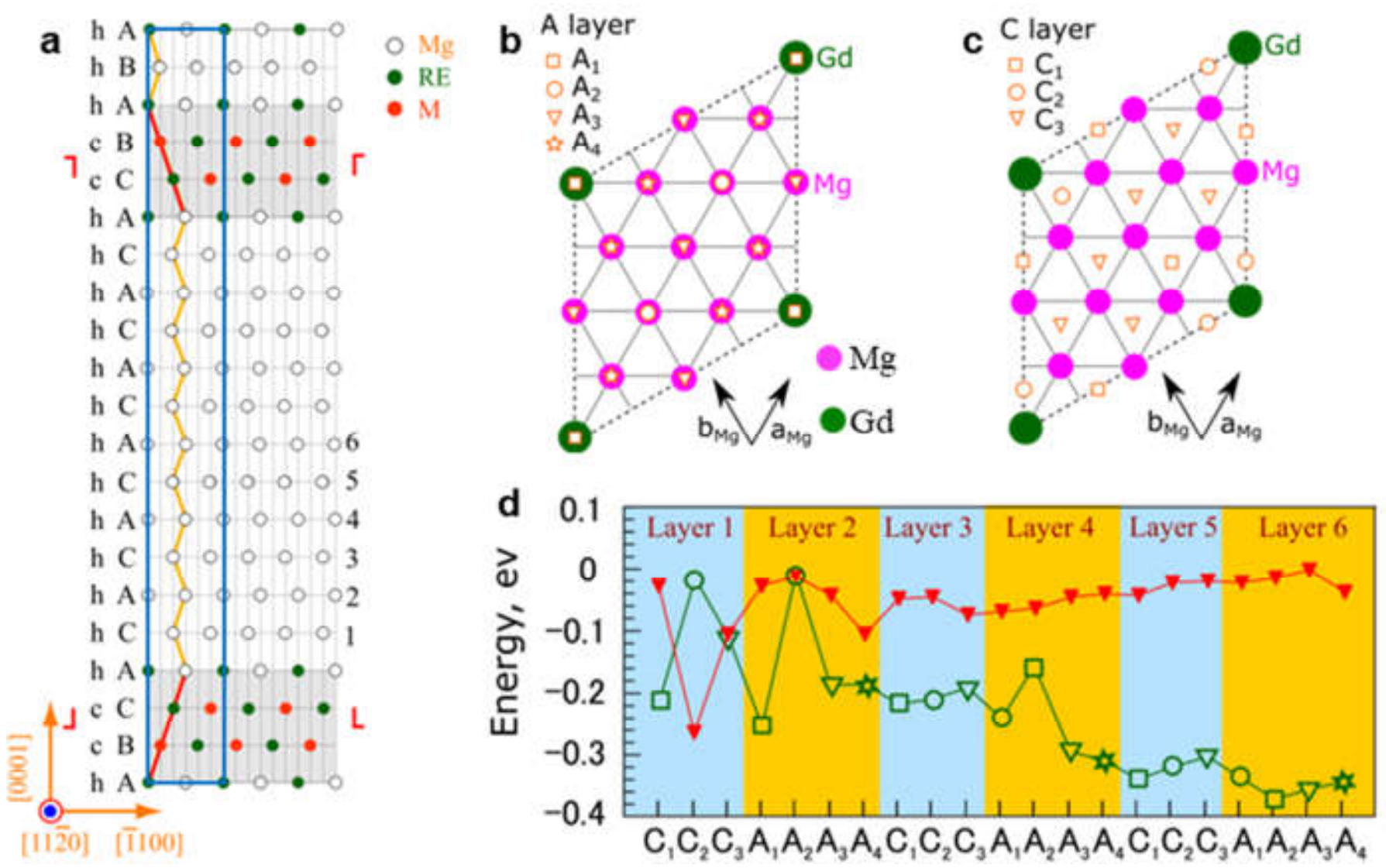

Figure 4. The total energy for the interaction between different solute atoms and SU. a) Schematic diagram of the computational model, the Mg atoms in Layer 1 to 6 will be substituted by solute atoms Al or Gd, b) four inequivalent sites in stacking layer A viewed along from $[0001]_{\alpha}$ direction, c) three inequivalent sites in stacking layer C, d) the energy for Al and Gd in different stacking layers and different positions. For simplicity, the total energy has been subtracted by their maximum value for Al or Gd.